# MEASURING THE VOLATILITY OF THE POLITICAL AGENDA IN PUBLIC OPINION AND NEWS MEDIA

CHICO Q. CAMARGO *
PETER JOHN
HELEN Z. MARGETTS
SCOTT A. HALE

**Abstract** Recent election surprises, regime changes, and political shocks indicate that political agendas have become more fast-moving and volatile. The ability to measure the complex dynamics of agenda change and capture the nature and extent of volatility in political systems is therefore more crucial than ever before. This study proposes a definition and operationalization of volatility that combines insights from political science, communications, information theory, and computational techniques. The proposed measures of fractionalization and agenda change encompass the shifting salience of issues in the agenda as a whole and allow the study of agendas across different domains. We evaluate these metrics and compare them to other measures such as issue-level survival rates and the Pedersen Index, which uses public-opinion poll data to measure public agendas, as well as traditional media content to measure media agendas in the UK and Germany. We show how these measures complement existing approaches and could be employed in future agenda-setting research.

Chico Q. Camargo is a lecturer in the Department of Computer Science at the University of Exeter, Exeter, UK, and a research associate at the Oxford Internet Institute, University of Oxford, Oxford, UK. This research was completed while he was a postdoctoral researcher in data science at the University of Oxford, Oxford, UK. Peter John is a professor of public policy at King's College London, London, UK. Helen Z. Margetts is Professor of Society and the Internet at the Oxford Internet Institute, University of Oxford, Oxford, UK, and director of the Public Policy Programme at the Alan Turing Institute, London, UK. Scott A. Hale is an associate professor and senior research fellow at the Oxford Internet Institute, University of Oxford, Oxford, UK, a fellow at the Alan Turing Institute, London, UK, and director of research at Meedan, San Francisco, CA, USA. This work was supported by the Alan Turing Institute [EPSRC grant EP/N510129/1 to H.Z.M.] and the Volkswagen Foundation [VWF grant 92 136 to S.A.H.]. *Address correspondence to Chico Q. Camargo, College of Engineering, Mathematics and Physical Sciences, Harrison Building, Streatham Campus, University of Exeter, Exeter EX4 4QF, UK; email: f.camargo@exeter.ac.uk.

The political agenda is the set of issues that are the subject of decision-making and debate within a given political system at any one time (Kingdon and Stano 1984; Baumgartner and Jones 1993). It is an important concept, as the attention paid by different parts of the political system to different events and to new ideas ultimately shapes public policy change.[1] For instance, understanding shifts in public opinion allows policymakers to respond to changing societal needs and to calibrate their responses to exogenous events. Such changes in the political agenda might come as part of long-term incremental developments or social trends, or they might come in the form of punctuations (that is, short, sudden, and dramatic shifts in public attention after a long period of relative stasis). This particular pattern is characterized by the famous Baumgartner and Jones (1993) model of "punctuated equilibria" (see also True, Jones, and Baumgartner 2019; Jones and Baumgartner 2012; Lundgren, Squatrito, and Tallberg 2018), which portrays how for many policy issues, long periods of stability are punctuated by sharp bursts of rapid, unpredictable change. On a global scale, major events causing punctuations include the financial crisis of 2007–2008, which placed the economy at the center of attention, or the September 11, 2001, attacks in the United States, which brought terrorism to the forefront of the political agenda of many countries. Such dominance of particular issues following punctuations can exclude other issues and mask changing policy needs, which may explain the lack of resources directed at public health following 9/11. It is therefore important for both policymakers and researchers to be able to track and understand the small-scale incremental changes taking place during both punctuations and the periods of stasis that follow. The richer and more fine-grained the information that policymakers have about changing public attention, the better they can respond to both social change and exogenous events.

The most developed way to assess agenda change and to capture these complex dynamics is through the measurement of volatility, that is, a measure of the tendency of a system to change rapidly and unpredictably. There are multiple ways of defining and measuring volatility across and within disciplines, including the natural sciences and social sciences. In this paper, we first discuss differences between alternative definitions of volatility. Second, we propose a definition of volatility in agenda setting that comports with concepts of volatility used in political science as well as those used in political communication research and information theory, and operationalizes this definition by developing measures of agenda fractionalization and agenda churn that can capture the dynamics of the agenda-setting system as a whole.

1. We use the term "political agenda" to refer to the general agenda of politics as defined by Kingdon and Stano (1984) encompassing public opinion and media attention. We use public agenda to refer specifically to public opinion and media agenda to refer to the issues covered by traditional media.

The measures discussed here can be used widely across the study of political systems, for example, to assess the extent to which volatility is increasing empirically in multiple spheres, effectively allowing for public opinion (the public agenda) to be compared against media salience (the media agenda), electoral politics, and legislative attention. We argue that these measures complement existing approaches to capture the overall effects of the complex dynamics of gradual and rapid agenda change over time. Third, the metrics proposed here are compared to other measures of volatility using four empirical case studies, taking data over the last 70 years from public opinion polls and traditional media in the UK and Germany. Finally, we discuss how the measures of volatility in political attention proposed here, combined with measures of volatility in other domains, could be employed in future research in agenda setting.

## Disciplinary Approaches to Volatility

In political science, a core notion of volatility is found in the Pedersen Index, which measures the net change within an electoral party system resulting from individual vote transfers between parties from any one election to another (Pedersen 1979). The Pedersen Index is also sometimes disaggregated into Type A or replacement volatility, which captures volatility in party entry and exit, and Type B or electoral volatility, which measures the volatility among stable parties that contest multiple elections (Powell and Tucker 2014).

Political communication studies of voter volatility have also studied volatile behavioral variables, such as unstable party affiliation, openness to alternatives, anti-party attitude, and vote abstention, all of which aim at measuring how a voter might behave in a volatile manner (Bybee et al. 1981). In spite of a need for care in using the index so the results are meaningful and comparable, it still remains the measurement tool of choice for volatility in political behavior and by extension to other kinds of behavior (Casal Bértoa, Deegan-Krause, and Haughton 2017).

In the policy agendas literature, scholars have studied the frequency and magnitude of individual fluctuations or events on the agenda. In general, the distributions of issue agenda fluctuations are leptokurtic and heavy-tailed. They are leptokurtic as most changes in the importance of an issue are very small, leading to a concentration around the mean. On the other hand, rare events called "policy punctuations"—sudden but rare breaks in an agenda series (John and Bevan 2012)—lead to outliers and a heavy tail. The volatility perspective introduced in this paper offers a more global assessment of changes feeding across the system as a whole rather than focusing on the distribution of fluctuations.

In (political) communication research, the agenda has been assessed in terms of three concepts: capacity (the number of issues on the agenda at any one time), diversity or fractionalization (how evenly attention is spread across issues, often measured with entropy), and issue volatility (which is usually defined as the speed with which issues move on or off the agenda), following the seminal work of McCombs and Zhu (1995, p. 497), which measured these three characteristics of shifts in US public opinion on "the most important problem facing the country at a particular moment," finding increasing agenda diversity and issue-level volatility from the 1950s to the 1990s. Edy and Meirick (2018, 2019) updated that work, finding that in spite of greater agenda diversity, issue-level volatility had changed little during the period.

This study builds on the work cited to conceptualize agenda volatility as a system-level phenomenon, encompassing both the fractionalization or diversity of the agenda and its change. As a result, it proposes two composite measures that conceptualize volatility more widely than proposed by McCombs and Zhu (1995). The two measures discussed here—the effective number of issues (ENI) and the novelty (or divergence) of the agenda—encompass the shifting salience of issues in the agenda as a whole, rather than averaging over the survival rate of individual issues. These measures are compared to alternatives through data from public opinion polls and traditional media from the last 70 years in the UK and Germany. A topic-modeling algorithm is applied to identify salient issues in textual news media, allowing for a comparison of the volatility in the media agenda and the public agenda.

## Quantifying Agenda Volatility

This study proposes two quantitative measures of political volatility drawing from information theory. The first measures fractionalization or the number of issues receiving attention simultaneously, while the second measures the change in the agenda between two points in time.

## Measuring Agenda Fractionalization

To measure the number of issues receiving attention, we propose the effective number of issues $ENI_t$, defined in Equation 1 as a function of the distribution of attention to a set $X$ of $n$ issues. Below, $p_t(x_i)$ indicates the attention received by issue $x_i$ at time $t$, and $H_t(X)$ stands for Shannon entropy (Kumar, Kumar, and Kapur 1986).

$$ENI_t = 2^{H_t},\ where\ H_t(X) = -\sum_{i=1}^{n} p_t(x_i)\ log_2\ p_t(x_i) \quad (1)$$

Since the effective number of issues is defined as the exponential of the Shannon entropy, this measure simultaneously captures all details captured by entropy while being a direct analogue to the effective number of species in an ecosystem (Jost 2006). Its value ranges from zero to $n$, the total number of issues on an agenda or the total number of species in a system. Its interpretation is also the same as for Laakso and Taagepera's effective number of political parties, which similarly draws from ecological diversity indices to measure the number of parties and their relative sizes in the electorate (i.e., vote share) or in the legislature (i.e., parliamentary seats) to calculate the number of "hypothetical equal-size parties that would have the same total effect on fractionalization of the system as have the actual parties of unequal size" (Laakso and Taagepera 1979, p. 4).

The effective number of issues is a measure of the fractionalization of the agenda at a given point in time: in other words, of how the political attention (of the public or the media) is divided between multiple simultaneous issues, weighing issues according to how much attention they receive to calculate what number of evenly considered issues would produce the same level of fractionalization. This measure is highest when attention is evenly distributed across many issues and lowest when it is concentrated on a single issue.

Laakso and Taagepera compare Equation 1 to the inverse Simpson index, defined as $N_S = 1/(\sum_{i=1}^{n} p_i^2)$,and argue that while they show the same qualitative features, the former definition is connected to the physical and information-theoretic concept of entropy. Other entropy-based measures include normalized entropy, which varies between zero and one regardless of the number of issues on the agenda (Jost 2006), and has been used many times in agenda setting since McCombs and Zhu's classic work (Chaffee and Wilson 1977; Culbertson 1992; McCombs and Zhu 1995; Edy and Meirick 2018; Pinto et al. 2019). Other works have already pointed out that normalized entropy does not allow for comparisons between agendas of different sizes (Boydstun, Bevan, and Thomas 2014), and have thus used the nonnormalized entropy instead—$H_t(X)$ in Equation 1 (John and Jennings 2010; Jennings et al. 2011; Baumgartner, Breunig, and Grossman 2019), sometimes alongside measures of electoral volatility (Mortensen et al. 2011).

While nonnormalized entropy makes results somewhat comparable across agendas of different sizes, it has no straightforward interpretation: it is difficult to understand what it means for an agenda to have entropy equal to 1.44, or 2.85, for example. The effective number of issues addresses that question with the exponential transformation mentioned above. The consequence of this simple transformation is that the *ENI*, unlike Shannon entropy, can be directly compared to real numbers of issues in an agenda. In other words, it can be interpreted as the degree of fractionalization of an agenda, ranging from 1, when all or nearly all attention is given to a single issue, to $n$, where public attention is distributed uniformly across $n$ issues. One would expect the latter scenario to be unlikely, given the limits on our capacity for processing information—the famous "magical number" of seven plus or minus two, which suggests one might expect five to nine issues to receive most of the attention at any single point in time (Miller 1956; Edy and Meirick 2018). An agenda with an effective number of issues equal to five, for example, would be as fragmented as an agenda with five issues of equal importance, thus echoing Laakso and Taagepera's description of the effective number of political parties (Laakso and Taagepera 1979).

## Measuring Changes in the Agenda

Since the effective number of issues only provides a snapshot of an issue agenda at one point in time, it cannot measure how attention to different issues changes over time. For this second aspect of volatility, another measure is needed.

Two broad families of measures are used to measure changes in the agenda. Issue-level measures describe the behavior of individual issues and are then averaged across all issues in the agenda. Agenda-level or system-level measures compare agendas (distributions of attention over multiple issues) directly.

The average issue survival rate, presented in McCombs and Zhu's (1995) work and in Edy and Meirick (2018), is an issue-level measure of how long the importance of a single issue is likely to remain above a given threshold. This measure is averaged across all issues in the agenda to give a sense of the agenda's stability, but it does not directly assess how different the whole agenda is at different points in time. Moreover, as Edy and Meirick (2018) point out, the Kaplan-Meier survival estimator adapted by McCombs and Zhu was originally intended to examine the product of a series of survival probabilities, such as how long a patient would survive after diagnosis (Kaplan and Meier 1958). This survival approach is not appropriate for the kind of data studied in agenda setting, given that "issues, unlike patients, can come back after mortality" (Edy and Meirick 2018). Survival measures cannot capture sudden rises in the importance of an issue. This poses a problem, as our understanding of agenda volatility encompasses any change in an agenda—both sudden rises as well as sudden decreases.

More generally, single-issue measures only provide a partial picture of agenda stability. While Baumgartner and Jones's original focus is on identifying agenda punctuations, there are usually long periods between punctuations, in which an agenda might show varying levels of stability. To understand these periods, whole-agenda measures are necessary.

A simple system-level approach would be to apply measures of voter churn based on the Pedersen Index, defined as the total net change within an electoral party system resulting from individual vote transfers, divided by 2 (Pedersen 1979), as shown in Equation 2 below for $p_t$, which designates an agenda at time $t$. For example, for an agenda composed of 20 issues, if one issue went down 20 percent while other two issues went up by 10 percent each, the resulting Pedersen Index would be of 20 percent ($(20 + 10 + 10)/2 = 20$).

$$PI_t(X) = \sum_{i=1}^{n} | p_t(x_i) - p_{t-1}(x_i)| \tag{2}$$

The Pedersen Index has been used in multiple contexts over the last decade (López, Vilchez, and Miranda 2011; Froio 2015; Bevan and Greene 2016, 2018; Carammia, Princen, and Timmermans 2016; Porta 2016; Green-Pedersen, Mortensen, and So 2018). Similar measures include the Agenda Stability measure introduced by Mortensen et al. (2011), based on the issue overlap measure developed by Sigelman and Buell (2004). Both papers produce the same aggregate measure of the total agenda change, and are identical to the Pedersen Index, except for a change in sign.

Despite their common usage, the simplicity of Pedersen-like indices is also their limitation. For instance, for the agenda described above, if instead all 20 issues were to change 2 percent in importance from one month to another, they would yield the same Pedersen Index of 20 percent (since $(2 \times 20)/2 = 20$). This simple observation suggests that Pedersen-like indices might not be an appropriate measure of unpredictability or surprise, since small fluctuations in the importance of many issues might produce the same Pedersen Index as a single large fluctuation. Given how common such small fluctuations in issue importance are (producing the leptokurtic distribution discussed earlier), this is likely a meaningful limitation. In contrast, a useful measure of volatility in political attention should reflect the fact that large, sudden changes in the political agenda are less likely than many small fluctuations.

Even though measurement systems have developed in robustly capturing changes in the allocation of attention and in government activity (Jones 2016), still no measure widely used in agenda setting addresses the weaknesses of the survival rate method and of the Pedersen Index, that is, a metric that measures the whole of the agenda, captures both sudden rises and sudden drops in issue importance, and differentiates many small changes from a few large changes. The Kullback-Leibler (KL) divergence (Kullback and Leibler 1951) is an information-theoretic measure that meets these requirements and complements the effective number of issues, another information-theoretic measure.

We propose the KL divergence between the distribution of attention over issues at a given point in time $p_t$ and same distribution at a previous time $p_{t-1}$ as a whole-agenda measure of agenda change. It is possible to speak of the KL divergence of an agenda at different time intervals (e.g., between weeks or years). The KL divergence between $p_t$ and $p_{t-1}$ is presented in Equation 3.

$$D_{KL}(p_t|p_{t-1}) = \sum_{i=1}^{n} p_t(x_i) log_2\left(\frac{p_t(x_i)}{p_{t-1}(x_i)}\right) \quad (3)$$

Unlike the Pedersen Index, KL divergence weighs changes by their relative magnitude, as indicated by the $\log_2(p_t(x_i)/p_{t-1}(x_i))$ term. Because of this weighting, relatively small fluctuations in an agenda result in a small KL divergence, whereas large fluctuations result in a large KL divergence.

Another name for KL divergence is relative entropy, as it is also an information-theoretic measurement of surprise, novelty, or information gain: If $p$ and $q$ are two probability distributions, $D_{KL}(p|q)$ is a measure of how much information is gained from $p$ assuming that $q$ is known. If they are the same, information gain is zero. It is important to differentiate the KL divergence from the Jensen–Shannon divergence, a symmetric measure derived from the KL divergence that has also seen application in the social sciences (DeDeo et al. 2013; Klingenstein, Hitchcock, and DeDeo 2014), including in agenda setting (Pinto et al. 2019), as a measure of the distance between two agendas. The Jensen–Shannon divergence, however, does not correspond to surprise or information gain. This is because it is a symmetrized measurement of the total difference between $p$, $q$, and the average of both distributions (Dagan, Lee, and Pereira 1997). This symmetry makes the Jensen–Shannon divergence a useful distance metric, but not a good measure of novelty, or surprise.

Apart from its many uses in physics and machine learning, the KL divergence has also been used to measure cognitive surprise (Itti and Baldi 2009), changes in semantic content (Barron et al. 2018), and information-seeking behavior (Murdock, Allen, and DeDeo 2017). It has also been used to study agenda divergence in social media (Park et al. 2013). As the KL divergence changes over time, lower values for $D_{KL}$ indicate periods of little change (relative stability in the agenda), while an increase in $D_{KL}$ indicates more instability, that is, a rise in volatility.

## Empirical Case Studies

The measures presented here were tested in four case studies, representing the combinations of two types of agendas, namely public opinion polls and news articles, with two countries, namely Germany and the UK. The reason for choosing such sources of agendas is that both environments show little friction and so may be subject to volatility and that both can be measured over a long period of time. As for the countries, the UK and Germany meet two different kinds of selection criteria. First, they represent different forms of democracy. The UK political system, which is centralized and lacks veto players, has been associated with receptivity to policy innovations and also liability to policy disasters (Dunleavy 1995; King and Crewe 2013), though the empirical record for these differences is less clear (Gray and Hart 1998). Germany has experienced relative stability since 1945 through coalition government, has a complex system of decentralization and intergovernmental relations, and usually experiences an incremental pattern of policymaking (Green and Paterson 2005). Second, both countries provide venues from which to observe changes over time—which is also verified for the public opinion polls and mass media in both countries.

Naturally, such a short list of data sources might raise questions about the applicability of these measures to new datasets. We note that the datasets considered here are not an exhaustive list of where the measures can be applied. Rather, they are simply the minimum necessary to illustrate the versatility of the measures discussed in this study: two public opinion polls collected in different countries over different time periods, and two news vehicles publishing in different languages. The formulas for the effective number of issues and the KL divergence require only a distribution of attention over multiple issues over time, and can therefore be applied easily to any dataset of this type.

## Public Opinion Polls

The public opinion poll datasets analyzed here are taken from attitudinal surveys that ask nationally representative samples of the British and German populations what they feel is the most important issue facing their countries on the day.

For the UK, the surveys used here are summarized as the Ipsos MORI long-term Most Important Issues Index, which corresponds to monthly surveys, each of a representative sample of approximately a thousand British adults, asking the open-ended question “What do you see as the most important issue facing Britain today?” and coding responses into different issue categories. The dataset considered here includes all monthly rounds of the Ipsos MORI survey since 1985 (earlier years were discarded due to most issues not being encoded in the first years of the survey).

The analogous dataset used here for Germany is the German Longitudinal Election Study (GLES), collected every three months. Although GLES is an electoral study, this investigation only considers its long-term cross-sectional surveys, which have been carried out at regular intervals since 2009, with stratified samples of German residents according to age, gender, and education. This study uses hand-coded responses to the free-response question "What is, in your opinion, currently the most important political problem in Germany? Please only state the most important problem," taken from 2010 onward. This long-term tracking produces data in a format comparable to the Ipsos MORI long-term Most Important Issues Index.

A heatmap showing UK public attention to different issues by month from 1985 to 2016 is shown in figure 1. In this heatmap, darker shades indicate higher importance of issues over time, and white rectangles represent issues that were absent from the survey during certain periods. In the figure, black annotations highlight events that drove attention to specific issues, such as the financial crisis of 2007–2008 and the September 11, 2001, attacks in the United States.

To examine shifts in public attention, the distribution of attention to issues over time was computed. Figure 2 shows the effective number of issues for each year in the British opinion polls at the top, and the same quantity for each quarter in the German public opinion polls at the bottom. The values are calculated following Equation 1. For the UK polls, only the top 20 issues with the most attention over the whole time span of the dataset were considered, and issues introduced in the UK polls after 1990 and in the Germany polls after 2012 were removed from the analysis. This results in only a slight variation in the number of total issues (see Supplementary Material figure 9). The total number of issues is stable at 16 after 1990 for the British polls and stays between 15 and 16 after 2011 for the German polls. This stability, however, is not observed in the effective number of issues: For the UK polls (figure 2, top), the ENI grows slowly and is marked by oscillations around a mean that seems to move slowly from 9 issues in 1993 to 11 or 12 issues in 2015. The German polls (figure 2, bottom) also show a slow growth from 9 issues in 2010 to around 12 issues in 2014, followed by a steep decrease in 2015. This decrease is due to the *Migration and Integration* issue becoming the dominant issue at the time.

While the effective number of issues in the public opinion polls shows clear trends for both countries, the KL divergence or novelty does not show any significant trend over time. Figure 3 (top) shows the novelty in the British polls, sampled every quarter to match the German polls. The plot shows a series of peaks between 1985 and 1990, due to the introduction of new issues in that period—evidently a great source of novelty in the agenda. Right after 1990, the plot shows a peak of a similar height that does not correspond to the introduction of any issues to the agenda. Instead, it represents a sudden shift in attention shortly after the IRA attacks in 1990–1991. Another peak appears in 2001, after the September 11 attacks in the United States. Since those periods correspond to a sharp increase in the *Defense/*

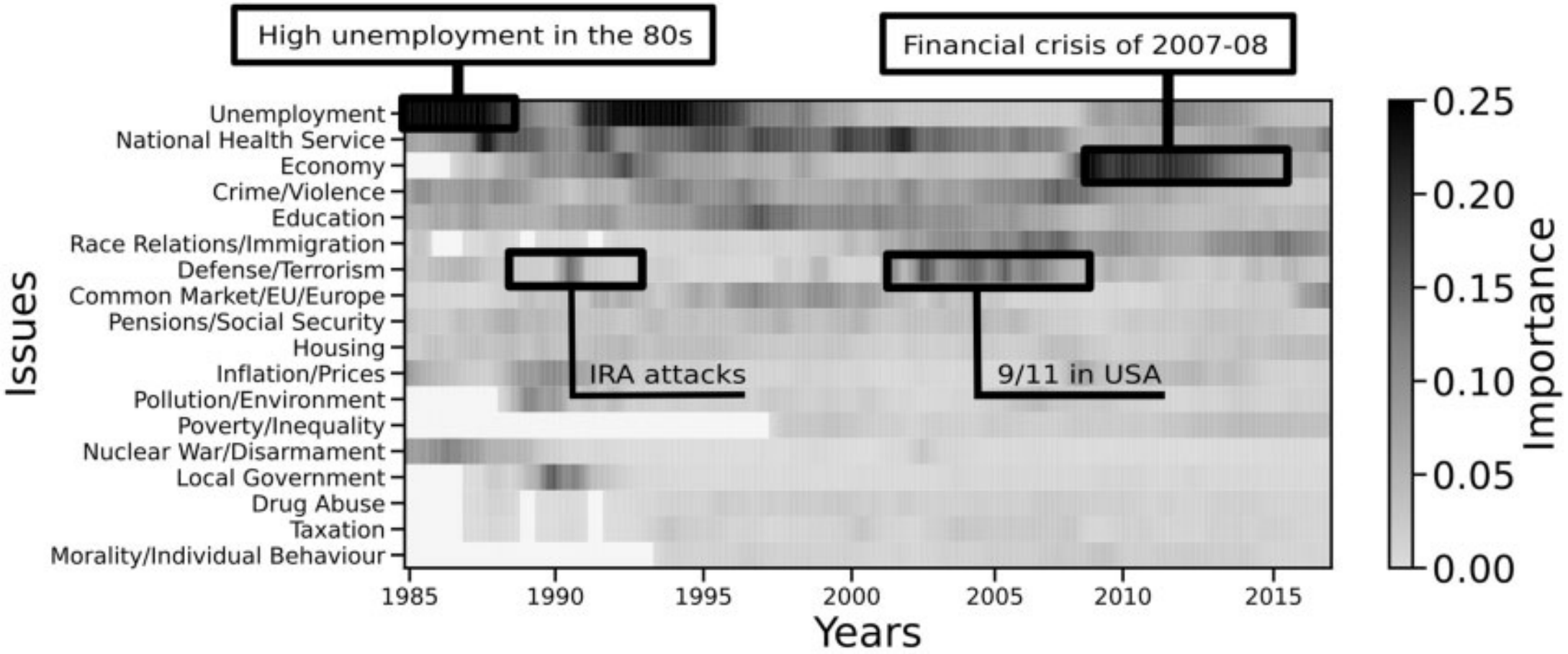

**Figure 1. A heatmap showing public attention to policy issues by month from 1985 to 2016 as reported by representative surveys of the UK population, collected by Ipsos MORI.** Darker shades indicate that a large fraction of the population indicated issues such as unemployment and the National Health Service as important. White areas indicate issues that were not included during a specific survey, and black boxes indicate specific events at the time of an increase in the attention to specific issues.

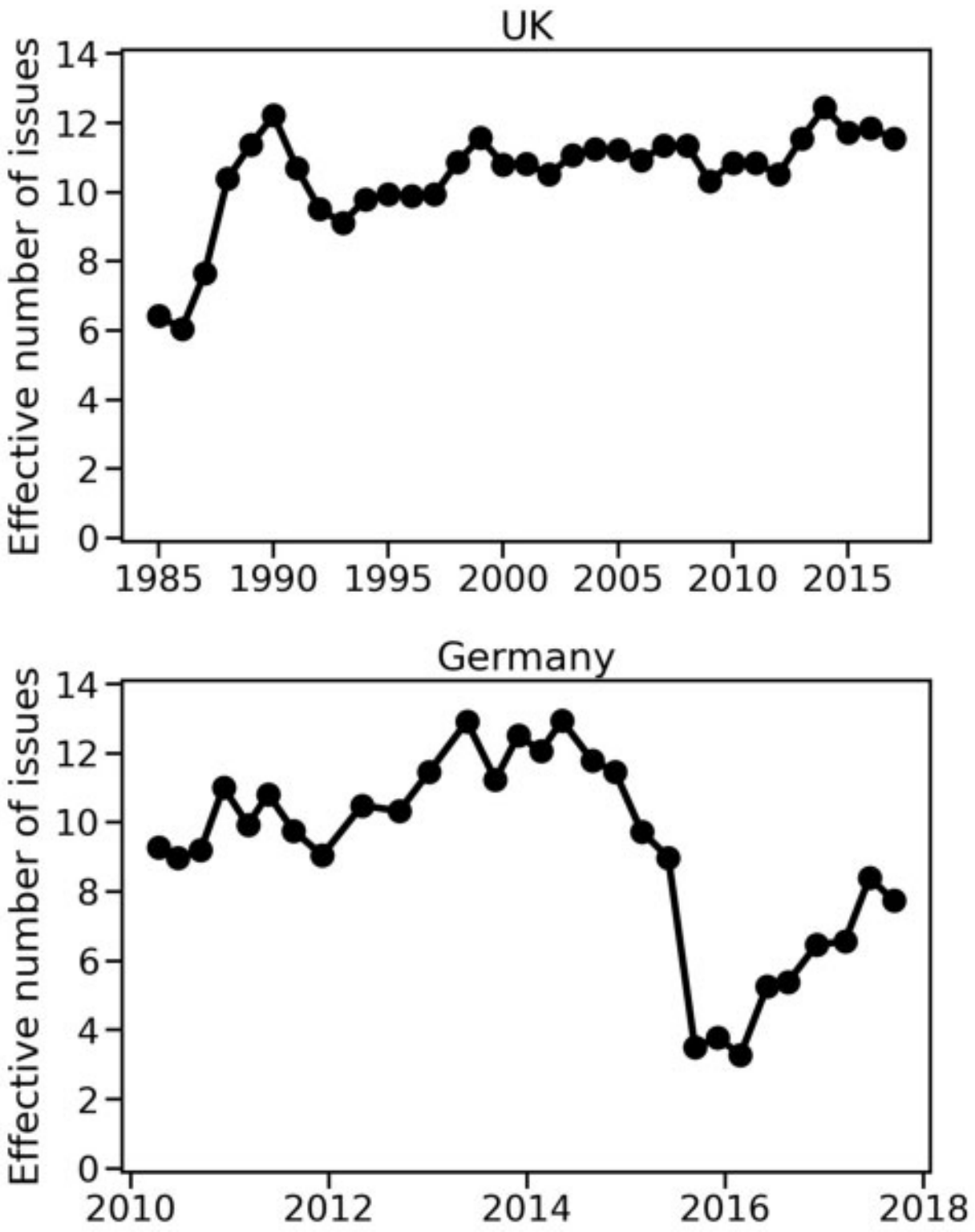

**Figure 2. Effective number of issues in UK and Germany public opinion polls.** Both panels show the change in the effective number of issues in the responses in the British public opinion survey data (top) and in the German opinion polls (bottom). Note that the time scales for the data sources differ markedly.

*Terrorism* issue (see figures 1 and 4), it is reasonable to attribute these peaks in novelty to the sudden relevance of this issue in the UK public agenda.

In contrast to the peaks corresponding to terrorist attacks, the 2008 financial crisis appears as a small bump in the time series in figure 3 (top), representing a much smaller increase in novelty than the peaks in 1990–1991 and 2001. This reflects the gradual growth in the importance of the *Economy* issue in the agenda, which took two years from 2007 to become the most important issue in 2009, as shown in figure 4. This growth in importance was slow in comparison to the sudden jumps in importance of the *Defense/ Terrorism* issue, shown in figure 4.

Figure 3 (bottom) shows the novelty for the German polls from 2011 to 2018. It has peaks in 2011, 2012, and 2016, but no sustained trend otherwise. The first two peaks coincide with a sharp growth followed by a sharp decline in the *Currency and Euro* issue. The peak in novelty in 2016 coincides with the growth of the *Migration and Integration* issue, which, as mentioned earlier, captured such a large share of attention that the number of effective issues also decreased at the time.

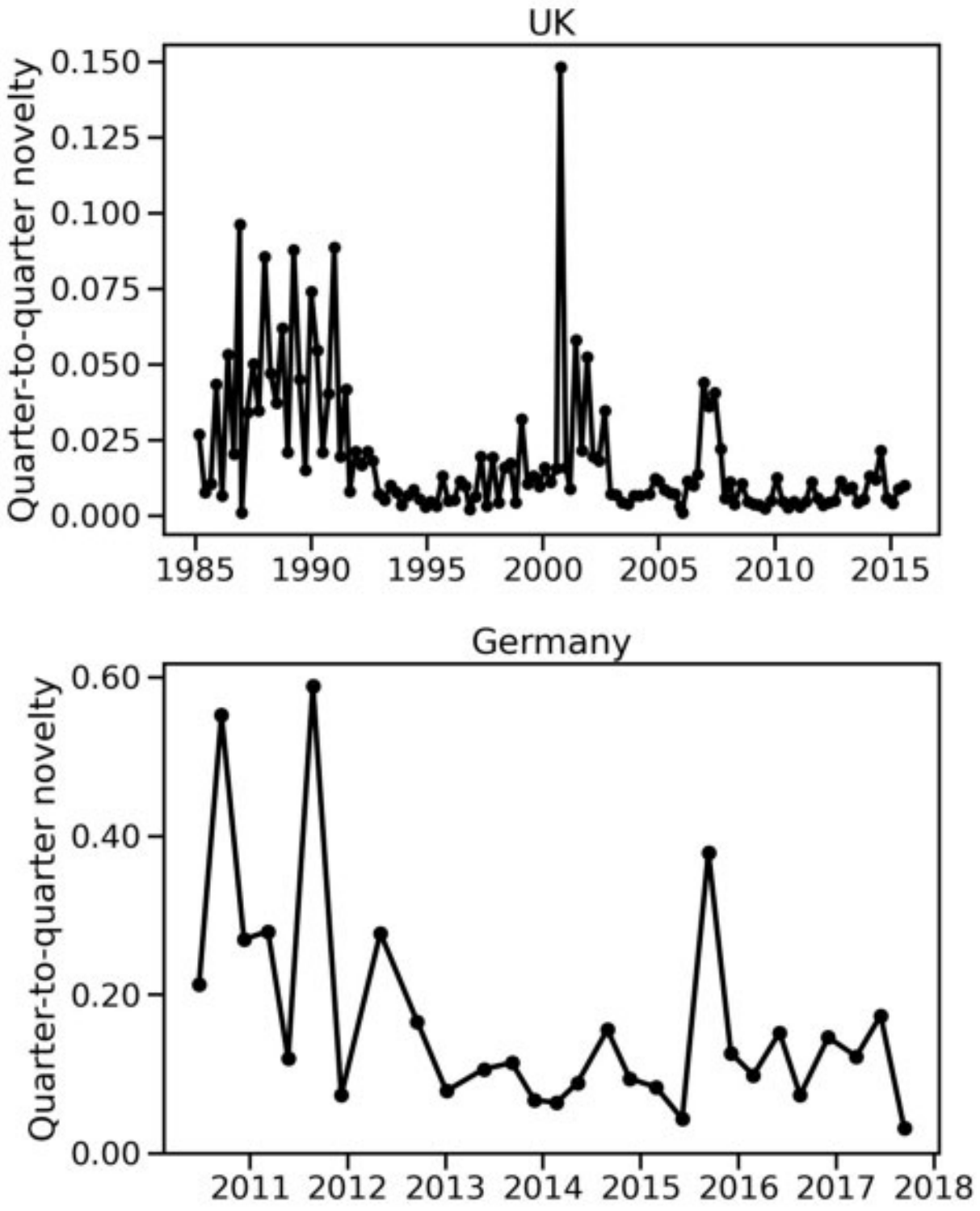

**Figure 3. Novelty in the issue agenda over time.** Both panels show novelty (KL divergence) measured for the distribution of attention to different issues over time, in the British monthly public opinion survey data (top), sampled every quarter, and in the German quarterly public opinion survey data (bottom). Higher values indicate sudden shifts in the public attention to policy issues, as well as the introduction of new issues to the policy agenda between 1985 and 1990. The German public opinion polls show higher novelty values, suggesting higher levels of unpredictability.

The German polls also show higher novelty than the corresponding UK polls shown at the top. At first sight, this result suggests that German public opinion might show more variation over time than British public opinion. Since figure 3 (top) was produced by subsampling the British polls quarterly, to match the frequency of German polls, and the total number of issues considered in each poll is approximately the same number (15–16 issues, as shown in Supplementary Material figure 9), any difference between countries is likely to come from another source.

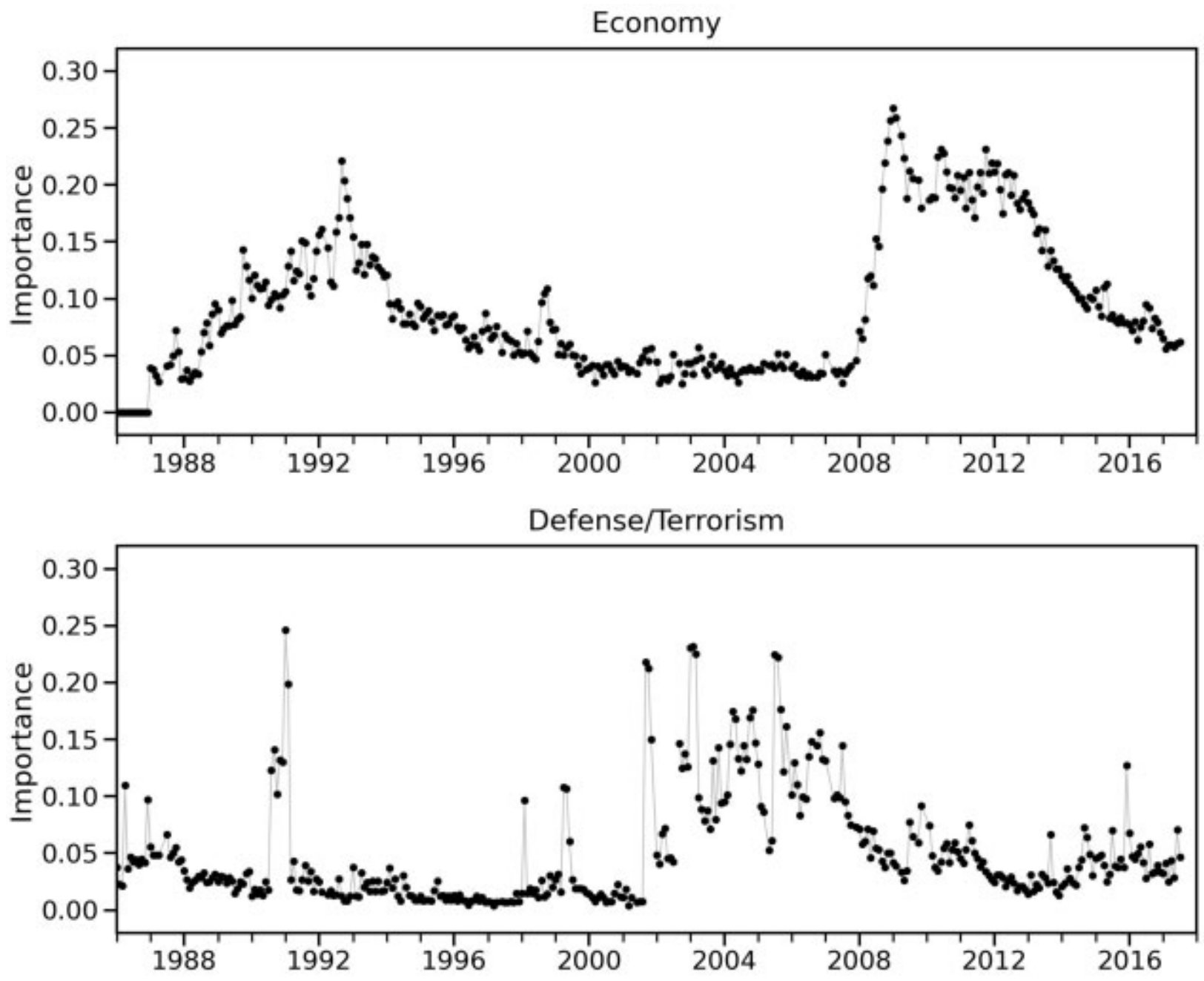

**Figure 4. Normalized importance of the issues *Economy* (top) and *Defense/Terrorism* (bottom) for the UK monthly public opinion polls.** While the salience of *Economy* rises gradually between 2007 and 2009, *Defense/Terrorism* shows a much more erratic behavior, characterized by short bursts of attention to the issue.

In addition to capturing policy punctuations in more detail, the KL divergence produces fewer spurious fluctuations than the Pedersen Index when applied to the same data. As the Pedersen Index yields high numbers in the presence of large amounts of small fluctuations, it shows a pattern of oscillations around a mean (figure 5). By contrast, KL divergence shows a value of approximately zero for most of the duration of the dataset, with large spikes when surprising shifts in the agenda occur, such as the new issues introduced by MORI during 1986–1990 or the September 11 terrorist attack mentioned above (figure 5).

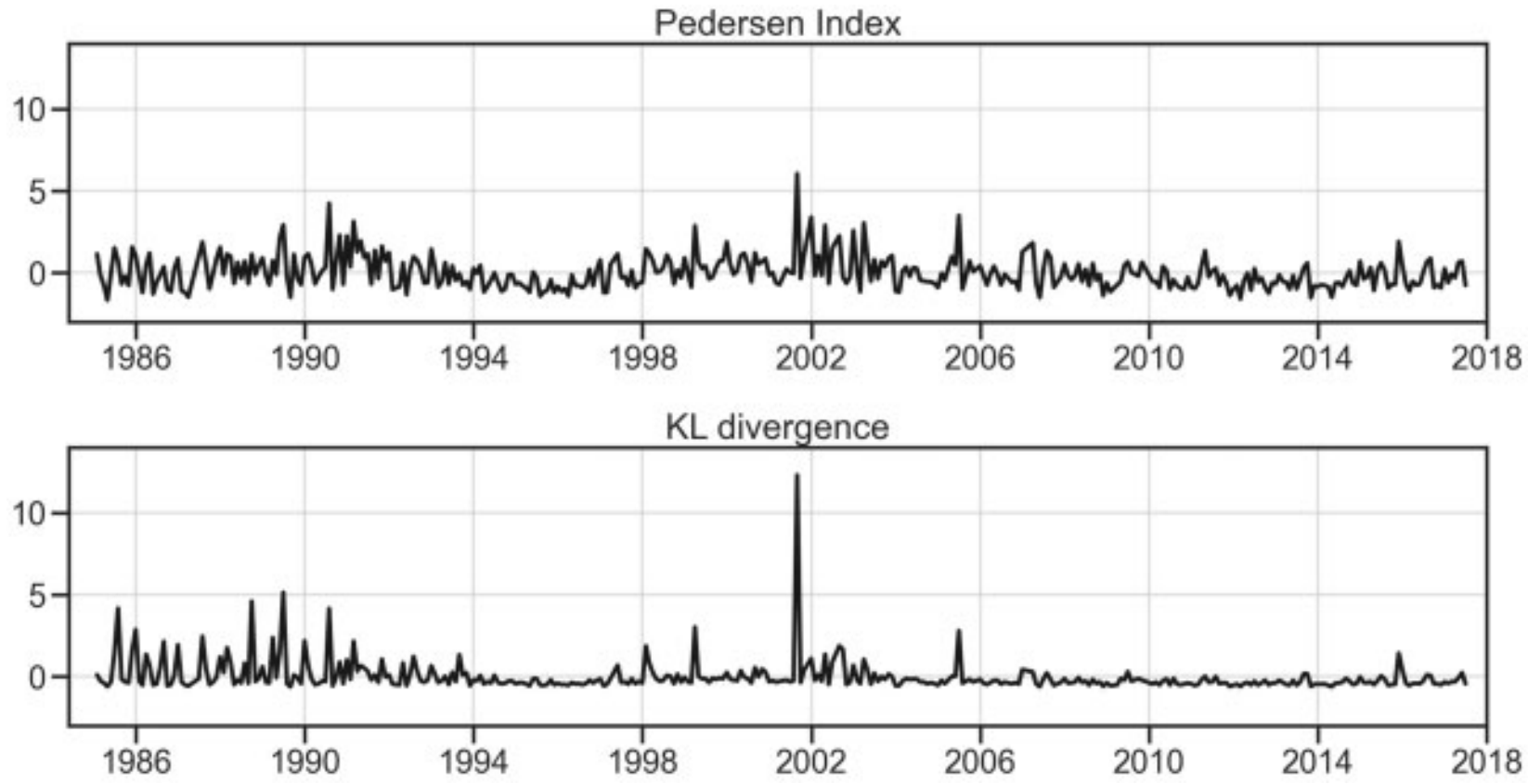

**Figure 5. KL divergence produces fewer spurious fluctuations than the Pedersen Index.** The top and bottom lines represent the month-to-month volatility in the agenda presented by Ipsos MORI UK opinion polls, as measured by the Pedersen Index and by the KL divergence, respectively, plotting both measures standardized around their means. As the Pedersen Index measures any change in the agenda, including large amounts of small, predictable changes, it shows a constant pattern of oscillations around its mean. In contrast, the KL divergence shows a value of approximately zero for most of the duration of the dataset, with spikes when surprising changes in the agenda occur, such as new issues introduced by Ipsos MORI between 1986 and 1990 and sudden shifts such as the September 11 terrorist attack in the United States.

## Traditional Media

For the British and German media agendas, topic modeling was used to reveal patterns of language use over time and uncover agenda themes. Topic models are widely used tools in large-scale text analysis, and can be used to estimate the semantic content of large collections of documents. For example, topic modeling has been used to analyze congressional speeches (Hillard, Purpura, and Wilkerson 2008; Quinn et al. 2010; Herzog, John, and Mikhaylov 2018), financial news (Curme et al. 2017), common issues raised in petitions (Vidgen and Yasseri 2019), and along with KL divergence, to measure the novelty of parliamentary speeches during the French Revolution (Barron et al. 2018).

The topic models representing the media agenda are produced using latent Dirichlet allocation (LDA) (Blei, Ng, and Jordan 2003), which assigns every article to a weighted combination of "topics," that is, co-occurring word patterns, and finds topics strongly marked by words related to political agenda issues, such as the national economy, unemployment, or war.[2] The LDA method allows for a variable number of topics, representing the level of coarse-graining of the text data. The results presented here consider 50 topics, although qualitatively similar results are found for numbers in the range from 10 to 90 topics, which includes the range of 15–20 topics present in the public opinion polls (see Supplementary Material).

The reason LDA topic modeling is particularly suited for this task is that the distribution of topics can be considered analogous to a distribution of attention to different issues. This analogy makes it straightforward to measure the effective number of issues and the KL divergence between different editions of the same newspaper or magazine. This can be done not by comparing the absolute values for these measures, as they are influenced by the number of topics chosen, but rather by comparing these measures across different data sources.

Figure 6 shows the effective number of issues measured for the UK-based daily newspaper *The Times* and the Germany-based weekly news magazine *Der Spiegel*. To allow for a better comparison between both sources, the dataset studied here considers one day a week of content from *The Times* (namely, Wednesday), starting from 1947, the year when *Der Spiegel* was founded. All articles for *Der Spiegel* from 1947 to 2016 were obtained directly from the news magazine's website.[3] Content for *The Times* was obtained from The Times Digital Archive, which contains digitized editions of the newspaper from 1785 to 2012 (Gale 2012). The Times Digital Archive includes a variety of sections. For this analysis, articles in "Arts and Sports" and "People," while keeping articles in "News," "Business News," and "Opinion and Editorial," were discarded.[4]

Examining the contents of the entire paper, rather than just the first or first few pages, is a direct consequence of the focus of this study: rather than studying only the most important issues in the newspaper, we track the distribution of attention across all issues discussed by the newspaper. This echoes the desire to measure the system-level change in the agenda over time, rather than only looking at punctuations or at its most important issues.

Both news publications show wide variation in the effective number of issues per week (see figure 6), showing overall growth from approximately 1955 to the mid-1990s, followed by a slow downturn, and a high-variance period in the early 2000s in *The Times*, and a sharp decline in the first half of the 2010s in *Der Spiegel*.

2. See Supplementary Material for example topics from one run.
3. One edition that was entirely in English, rather than German, was excluded.
4. Retaining only articles classified as "News" produces similar results.

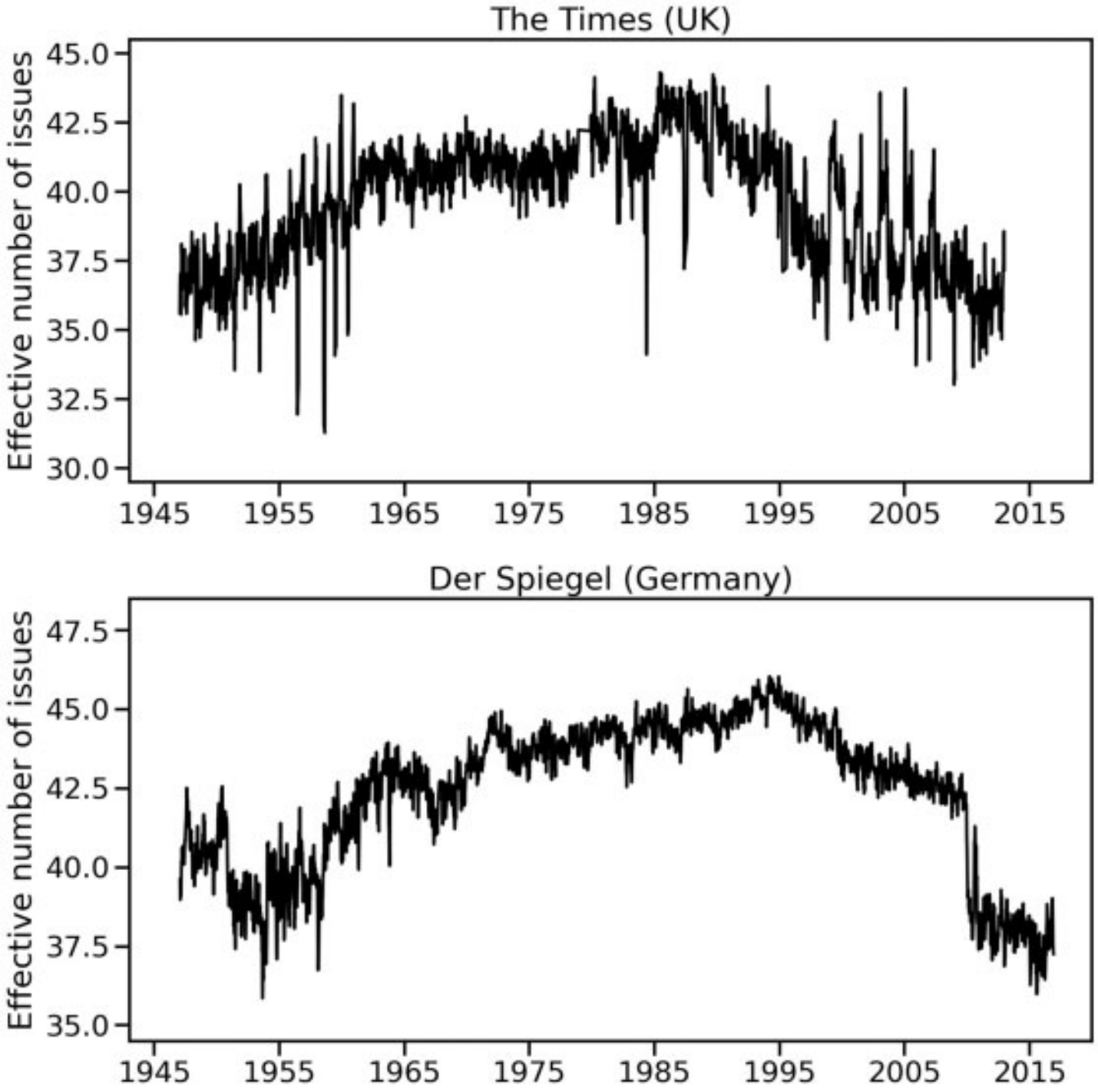

**Figure 6. Effective number of issues per week in UK and Germany media.** The two plots show the change in the effective number of issues over time, for the British newspaper *The Times* (top), and for the German news magazine *Der Spiegel* (bottom)**.**

As discussed above, the effective number of issues measures the fractionalization of issues present in an agenda. Under the lens of KL divergence, *The Times* and *Der Spiegel* show different behavior from each other. As shown in figure 7 (top), the month-to-month novelty in *The Times* is approximately constant from 1947 to 2012, suggesting that its monthly churn of attention to different issues is relatively constant over time. For *Der Spiegel*, at the bottom of figure 7, the month-to-month novelty shows a slow increase, which is sharper in the 1960–1970 decade. Novelty in *Der Spiegel* also increases in early 2010, simultaneously to the sudden decrease in the effective number of issues shown in figure 6. These trends in novelty are not affected by the number of topics, for both *Der Spiegel* and *The Times*.

The news media data also provide a good illustration of the difference between the KL divergence and other measures. The difference between the KL divergence and averaged issue survival rates is particularly evident when large shifts in the agenda do not consist in the decay of any particular issue, but rather in the sudden rise of a group of issues. This is true for all major punctuations in *The Times* (figure 8), which correspond to moments when groups of issues suddenly rose in importance in the agenda—which results in a high KL divergence, but which are not well captured by issue survival rates nor by the Pedersen Index.

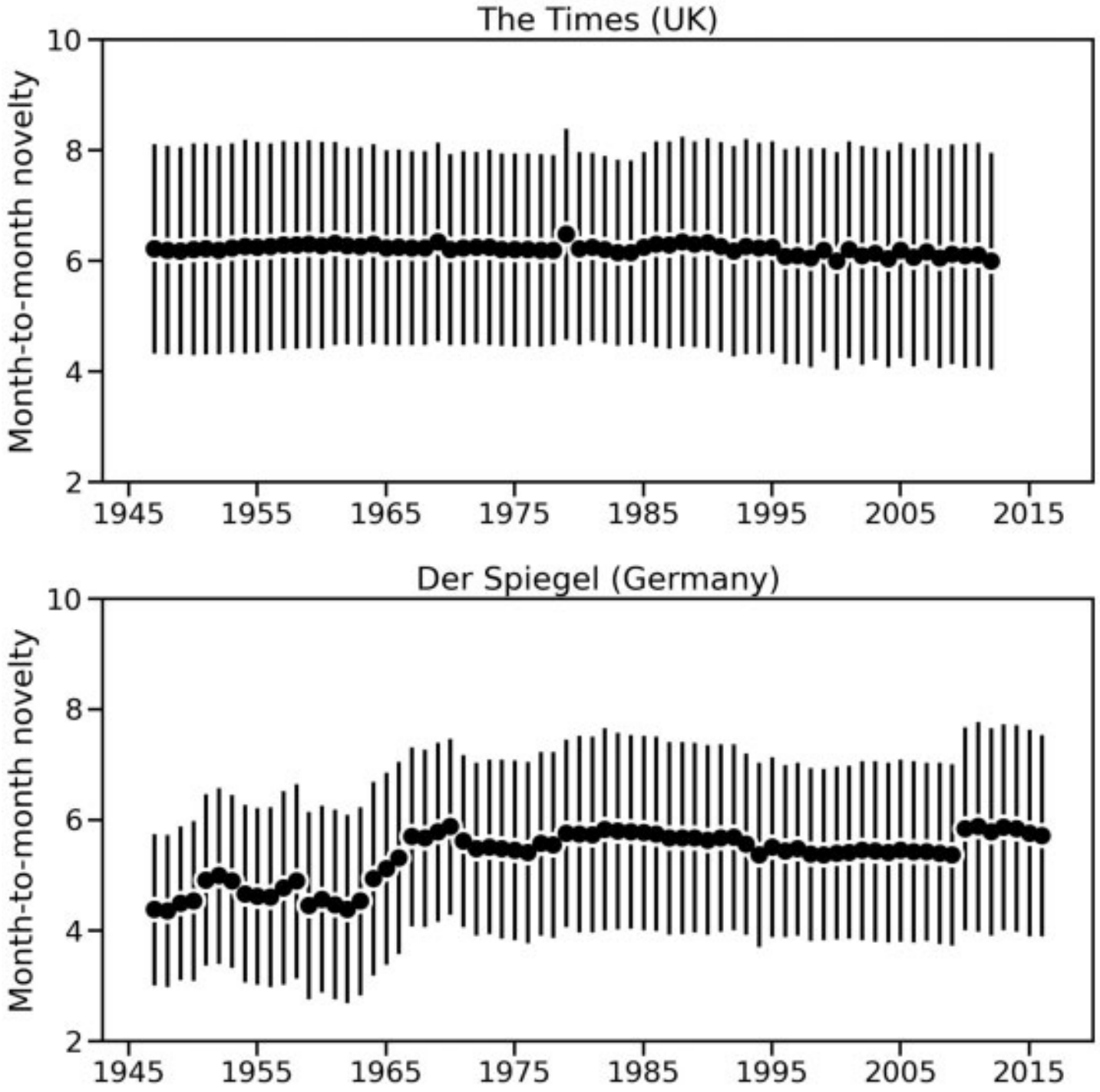

**Figure 7. Novelty in the media agenda.** Time series plots showing the novelty (KL divergence) measured for the distribution of attention to different issues over time, in the British newspaper *The Times* (top), and in the German news magazine *Der Spiegel* (bottom). Higher values indicate more sudden shifts in the attention dedicated to policy issues in news media.

## Conclusions

This paper conceptualizes the idea of volatility as it applies to the whole of the political agenda, bringing together concepts, models, and frameworks from political science, communication research, information theory, and computational techniques, to complement existing measures of political attention across media sources and public opinion. Following a review of measures of volatility in other domains, we operationalize the concept with two measures. The first, the effective number of issues, measures the fractionalization of the agenda, or how many issues are being considered simultaneously, and the second, the KL divergence or novelty, measures how quickly attention shifts from one set of issues to another over time. These measures complement existing scholarship in two key ways. First, the effective number of issues provides a measure of agenda fractionalization that allows for the same intuitive interpretation as Laakso and Taagepera's effective number of parties and allows for comparisons across agendas with different numbers of issues. This stands in contrast to the normalized measure of diversity provided by McCombs and Zhu (1995), where entropy is normalized to always fall between zero and one regardless of how many issues are on an agenda. Second, KL divergence, our measure of agenda change, is a measure for surprise tested in many other disciplines. As it measures change at the level of the whole agenda, it differs from the average issue survival indices often used in the literature (McCombs and Zhu 1995; Edy and Meirick 2018). The KL divergence also separates large policy punctuations from many small changes in a way that the Pedersen Index does not. The fact that KL divergence can also be interpreted as information gain, novelty, or surprise provides a natural interpretation for the measure: it quantifies how surprising an agenda at a given point in time is when compared to the same agenda at another point in time.

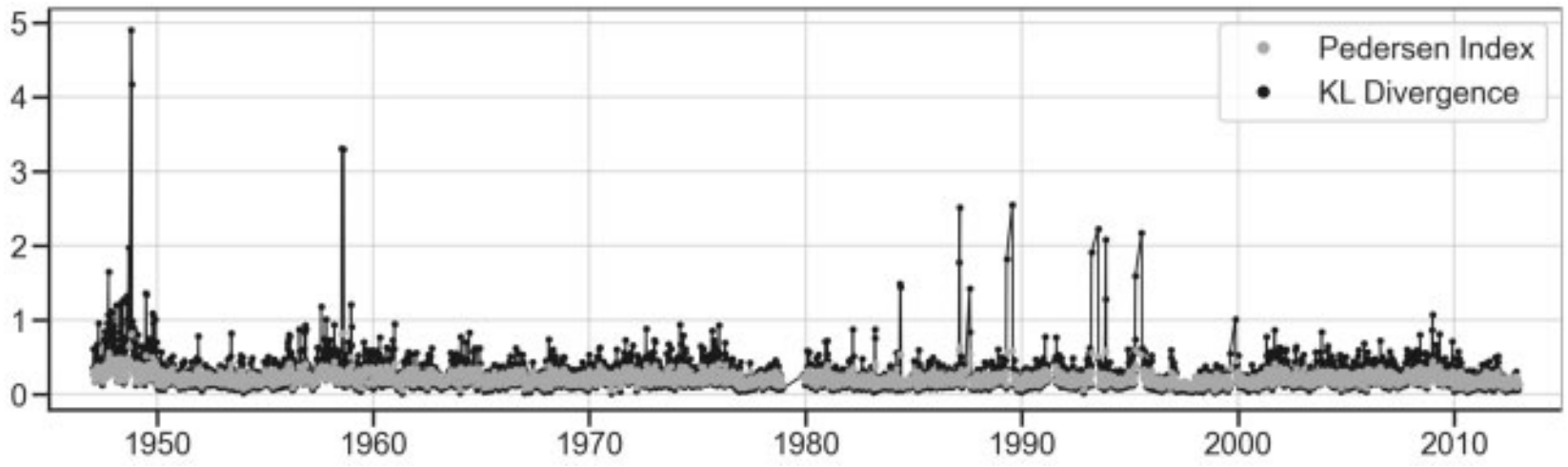

**Figure 8. KL divergence captures policy punctuations that the Pedersen Index fails to capture.** In this figure, points represent measures of week-to-week volatility in the agenda presented by *The Times*, during all weeks ranging from 1947 to 2012, excluding the period from December 1978 to November 1979, when the paper was shut down due to a strike. All major agenda punctuations shown during the period are captured by the KL divergence measure, shown in black, but not as much by the Pedersen Index, shown in gray. The effect remains even when both measures are standardized.

Both measures are then used to assess the volatility of public opinion polls and news media data in the United Kingdom and Germany. Topic modeling is used to translate news articles into distributions of attention to different issues, thus making it possible to measure the effective number of issues active in the news media, and also to calculate the month-to-month change in the news media agenda as a whole.

Not only does the effective number of issues in each dataset vary over time, but the estimate that the public would consider of 5 to 9 issues at a time (Miller 1956) is also not met. Rather, the data show larger numbers of issues under consideration in the public opinion polls and in the news media, even when news sources are modeled using only 10 possible topics. It is worth noting the excellent point made by Edy and Meirick (2018) that the limits of an individual's working memory do not necessarily limit the scope of the public agenda. We hypothesize that in today's digital environment, with news sources spread across social media platforms, one would expect collective public attention to be able to keep even more issues simultaneously on the public agenda than in a conventional media environment. Figure 2 seems to point in that direction, with the effective number of issues often above 9 for both countries, but a more thorough comparative study will be necessary to test this hypothesis. Nevertheless, the possibility of a varying effective number of issues implies that measures of agenda fractionalization should not be normalized between zero and one, but rather should allow for a changing number of issues in the agenda.

The results here show that peaks in the KL divergence correspond to policy punctuations, that is, large shifts in attention (Baumgartner and Jones 1993). These generally correspond to issues that capture a large amount of interest suddenly. Notably, events such as the financial crisis appear as much smaller fluctuations in the time series, representing an important type of policy change characterized by gradual shifts in attention rather than by large punctuations. Aside from these punctuations, there is no clear trend in the KL divergence across the opinion polls and news media data.

For the news media, it is worth noting that both newspapers went through many changes from 1947 to 2016, many affecting content and length of publications. Both *Der Spiegel* and *The Times* changed formats multiple times within this time range, alternating between broadsheet and tabloid formats, launching online and regional editions, and being affected by political changes such as the reunification of Germany in 1990. Naturally, any changes in the effective number of issues due to factors such as these cannot be determined from volatility measures alone. One confound that can be ruled out, however, is a change in the number of articles per month: as shown in Supplementary Material figure 11, the total number of articles published per week decreases between 2000 and 2010; however, the corresponding drop in the effective number of issues is larger than what would be expected for a null model controlling for the number of articles. This suggests that *Der Spiegel* was covering even fewer issues than expected when taking into account the number of articles during the period.

It is also important to note that the data presented in this paper do not capture recent historical developments involving UK and Germany, such as the political crisis in the UK following the 2016 EU membership referendum which, along with earlier events such as the demonstrations after the Arab Spring from 2011, have suggested an increase in system instability or volatility (Margetts et al. 2015), causing some leading political scientists to ask in both scholarly and media outlets "why is democracy so surprising?" (Gamble and Wright 2019; Runciman 2019a, 2019b) and others to argue that the rising use of social media is implicated in increasing volatility (Margetts et al. 2015). In contrast to this perception, the month-to-month shift in attention has been relatively constant. There have been a number of punctuations driven by issues such as terrorism and migration, but only in Germany has the effective number of issues shifted strongly within the time period studied. It may be that there is always a tendency for popular perception of volatility to run ahead of what can be measured, but further research is needed to test these hypotheses about the seemingly turbulent years following the events of 2016. For example, a multi-country comparison of internet and social media penetration and volatility in political attention could meaningfully contribute to the debate over the role played by the internet and social media in increasing the number of simultaneous issues receiving attention (Margetts et al. 2015; Edy and Meirick 2018).

Overall, the development of new ways to measure volatility in political attention opens exciting avenues for political science and agenda-setting research, outlining a future research program that can ascertain the extent and variation in volatility in political attention and investigate possible underlying causes in comparative contexts. The more rigorous operationalization of the concept of volatility in agenda setting presented here, done through information theory and the KL divergence, also raises further questions regarding the connection between novelty, changes in public concern, legislative action, and democratic outcomes. The measures that we have operationalized can be applied to both media attention and public opinion, as done here, but future research might apply the same measures to other domains within political systems, such as public expenditure and legislative attention, in terms of how much parliamentary time is spent on different issues. With an information-centric and multidisciplinary approach, our measures of volatility in political attention will enable further research in this important area.

## Data Availability Statement

REPLICATION DATA AND DOCUMENTATION are available at: https://github.com/euagendas/POLVOL.

## Supplementary Material

SUPPLEMENTARY MATERIAL may be found in the online version of this article: https://doi.org/10.1093/poq/nfab032.